# Structural flexibility dictates reactivity of single-atom catalysts

Jakub Planer[1], Dominik Hrůza[1], Tadeáš Lesovský[1], Ayesha Jabeen[1], Jan Čechal[1,2*], Zdeněk Jakub[1*]

[1] CEITEC – Central European Institute of Technology, Brno University of Technology, Purkyňova 123, 61200 Brno, Czechia

[2] Faculty of Mechanical Engineering, Brno University of Technology, Technická 2896/2, 616 69 Brno, Czechia

Correspondence to: cechal@fme.vutbr.cz, zdenek.jakub@ceitec.vutbr.cz

**Abstract**

Unravelling the origins of single-atom catalyst reactivity is a central challenge in heterogeneous catalysis research. A key question is whether the activity arises solely from atomic isolation or from distinct structural and electronic configurations of the single atoms. Here, we use precisely defined Fe–N$_3$ and Fe–N$_4$ model catalyst sites synthesized on an inert support to quantify the effect of coordination geometry on chemical reactivity. Both the Fe–N$_3$ and Fe–N$_4$ models have the same electronic configuration (high-spin Fe$^{2+}$ with S=2), and even their d-orbital occupancies and positions with respect to Fermi level are almost identical. Despite this electronic similarity, the adsorption energy of CO differs by more than 0.6 eV between the Fe–N$_3$ and Fe–N$_4$ sites, as indicated by density functional theory computations and confirmed by atomically-resolved scanning tunneling microscopy experiments. We trace this reactivity difference to the structural flexibility of the Fe-N$_3$ sites, which allows strengthening of the Fe 3d$_{xz/yz}$ - CO 2π* back-bonding by lifting the Fe atom from the -N$_3$ plane. These results demonstrate that coordination geometry plays a crucial role in defining the reactivity of single-atom catalysts, and that such effects cannot be predicted by analysis of the sites' electronic structures alone.

**Introduction**

Single-Atom Catalysts (SACs) featuring metal atoms stably embedded in carbon-based materials show high promise for many reactions including oxygen evolution/reduction,[1–4] hydrogen evolution,[3–5] methane conversion,[4,6] alkene hydrogenations[7,8] and hydroformylations.[3,9] The motivation to utilize "single-atoms" instead of larger clusters is obvious: the best catalysts are often precious metals, thus reducing the amount needed can greatly improve the economic viability of technologies that are currently too expensive to become widespread. Recent years saw an explosion of SAC synthesis and analysis methods, but rational design remains a challenge because the reactivity of SACs is not easy to predict. Noble metals like Au can become reactive when dispersed as single Au$_1$ atoms,[10–12] while traditional catalysts like Pt may show the opposite trend with their reactivity being decreased at the single-atom limit.[13–15] These variations complicate SAC design, but they also open realistic opportunities for developing SACs from abundant elements like Fe or Ni that could surpass the expensive catalysts of today.

The reactivity of SACs is equally dependent on the metal atom itself and its surroundings.[16–27] The local environment defines the crucial parameters like the coordination number of the metal atom, its oxidation state, d-orbital filling, spin configuration or accessibility to reactants. Modifying one of these parameters typically perturbs several others, making the search for clear reactivity trends a very complex task. Various "universal descriptors" of SAC reactivity were proposed in the literature,[28–32] many of which are derived from the so-called "d-band center model" developed on extended surfaces of metals and alloys.[33–35] However, experimental verification of the proposed reactivity trends is challenging, mainly because the reactivity differences between similar sites are difficult to ascertain on high-surface-area catalysts. The most reactive SAC sites are rarely the most abundant ones, thus they can be easily overlooked in bulk-averaged datasets.[36,37] It is now widely accepted that undercoordinated sites like metal-$N_3$ are likely to be more reactive than the more commonly found metal-$N_4$,[26,38–40] but direct experimental views on the fundamental reasons for the increased reactivity are missing.

To address the challenges outlined above, we developed a model platform in which a single reactivity-defining parameter can be intentionally varied, and the reactivity trends can be unambiguously ascertained using atomically-resolved experiments and computational modelling. Specifically, we synthesized 2D metal-organic frameworks (MOFs) containing perfectly defined Fe-$N_3$ and Fe-$N_4$ sites residing atop an inert graphene support. Then, we quantify the reactivity of these Fe-$N_3$/-$N_4$ sites towards carbon monoxide (CO), a crucially important molecule in both fundamental studies and applications. We use variable-temperature Scanning Tunneling Microscopy (STM) to directly image the CO molecules adsorbing and desorbing from the individual Fe-$N_x$ sites, and our temperature-dependent STM dataset allows us to quantify the CO adsorption energy. Our data show that CO is stable on the Fe-$N_3$ sites up to 180 K, but it does not adsorb on the Fe-$N_4$ sites even at temperatures as low as 115 K. This is consistent with Density Functional Theory (DFT) computations showing that the CO adsorption energy differs by 0.62 eV between the two systems. We demonstrate that the origin of the different reactivity lies in the structural flexibility of the Fe-$N_3$ sites, which allows significant lifting of the Fe atoms from the -$N_3$ plane. This structural change allows more efficient orbital hybridization which stabilizes the Fe-CO bond by 0.61 eV, primarily through the Fe $3d_{xz/yz}$-CO $2\pi^*$ back-bonding channel.

Overall, our experimental and computational approach identifies a direct link between structure and reactivity, and demonstrates that structural flexibility can play a decisive role in adsorption energetics. Thus, this parameter needs to be considered in any proposed universal descriptors of single-atom reactivity.

**Results and Discussion**

*Synthesis and analysis of Fe-N₃ and Fe-N₄ models*

We have chosen Fe-DCA (DCA = 9,10-dicyanoanthracene) and Fe-TCNQ (TCNQ = 7,7,8,8-tetracyanoquinodimethane) 2D Metal-Organic Frameworks as suitable models containing high surface density of well-defined Fe-N$_4$ and Fe-N$_3$ sites. Both systems were synthesized on a chemically inert graphene/Ir(111) support according to protocols described in Methods section. Figure 1 shows room temperature STM images, Low-Energy Electron Diffraction (LEED) patterns and DFT models of both model systems. Figure 1A-C reveals that Fe-DCA/graphene crystallizes in a honeycomb-kagomé structure with Fe$_2$(DCA)$_3$ stoichiometry, consistently with previous reports of 2D metal-DCA structures.[41–46] The Fe cations are coordinated to three cyano- groups, resulting in stable Fe-N$_3$ sites. This structure is present in two symmetry-equivalent domains on the graphene/Ir(111) support; these domains are rotated by ±12° from the high-symmetry directions of the support. This is well visible in the LEED pattern shown in Figure 1B, and in large-scale STM images presented in Figure S1. Importantly, DFT computations indicate that the weak interaction with the graphene support allows the Fe-N$_3$ site to remain planar (see the side view in Figure 1C); in contrast to similar systems placed on metal supports.[46]

Figure 1D-E shows a similar dataset acquired on the Fe-TCNQ/graphene/Ir model system. Fe-TCNQ 2D MOF crystallizes in a rectangular lattice with Fe$_1$(TCNQ)$_1$ stoichiometry, as shown in the STM image and DFT model presented in Figure 1D,E. The Fe cations are coordinated to four cyano- groups, resulting in stable Fe-N$_4$ sites. Previous literature indicates that the most stable Fe-TCNQ structure is physically corrugated due to the preference of the Fe cations to reside in tetrahedral coordination environment.[47] Such non-planar structure is also stable on the weakly-interacting graphene/Ir support, although its physical corrugation is slightly reduced by the van der Waals interaction with the support (shown in the side view in Figure 1E, details in reference [47]). The physical corrugation of the Fe-N$_4$ sites can be observed in room-temperature STM data acquired at specific tunneling conditions, as shown in Figure S2 in the SI. The Fe-TCNQ structure is present in 15 symmetry equivalent domains on the graphene/Ir support, as shown in the LEED pattern presented in Figure 1F.[48]

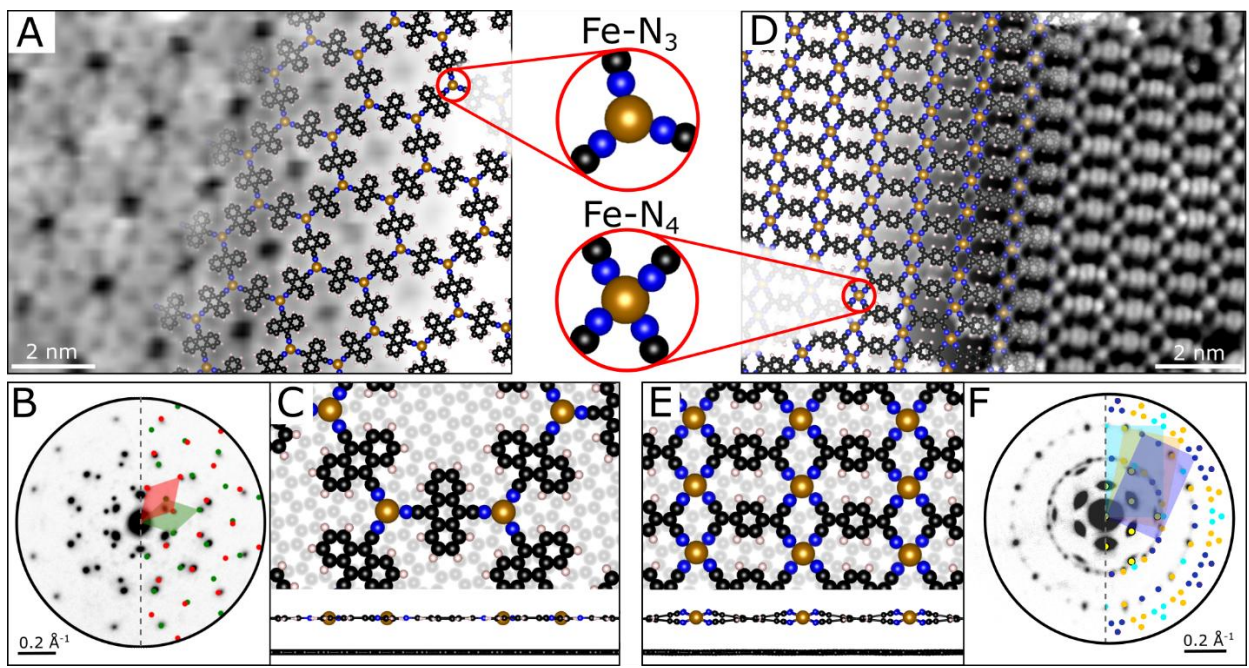

Figure 1: Structural characterization of Fe-DCA and Fe-TCNQ 2D MOFs on graphene by Scanning Tunneling Microscopy (STM), Low Energy Electron Diffraction (LEED) and Density Functional Theory computations (DFT). (A,B,C) The Fe-DCA structure accommodates Fe in 3-fold coordinated Fe-$N_3$ sites, as visible in the STM image and DFT model. The LEED pattern shows that the structure is present in two orientations on the graphene/Ir support. DFT model shown in panel C indicates that the interaction between graphene and Fe-DCA is weak; the Fe-$N_3$ sites remain planar. (D,E,F) The Fe-TCNQ features 4-fold coordinated Fe-$N_4$ sites. LEED pattern indicates that this structure is present in 15 orientations on the graphene/Ir surface. In DFT models, the carbon atoms are black, nitrogen atoms are blue, Fe atoms are brown and hydrogen atoms are white. The STM scanning parameters are $U_{sample}$=−0.6 V, $I_{bias}$=0.06 nA (A), and $U_{sample}$=1.2 V, $I_{bias}$=0.06 nA (B).

The Fe cations within Fe-DCA and Fe-TCNQ differ in their local coordination geometry, but they are remarkably similar in their electronic structure. Our DFT results shown in Figure 2 indicate that the Fe within Fe-DCA is in high-spin $d^6$ (S=2) $Fe^{2+}$ state, and this assignment is consistent with recent experimental characterization of a similar system by X-Ray Adsorption (XAS).[46] Figure 2A shows the electronic density of states (DOS) projected on the Fe 3d-orbitals within the Fe-DCA and Fe-TCNQ structures. The features observed in the DOS plots can be qualitatively rationalized by the crystal/ligand field theory: in Fe-DCA structure, the coordination geometry of Fe is trigonal planar, leading to degeneracy of symmetry-equivalent orbitals $d_{xy/(x2-y2)}$ as well as $d_{xz/yz}$. Our computational results indicate that these four Fe 3d orbitals are singly-occupied, while the $d_{z2}$ (which has minimal spatial overlap with the -CN groups of the DCA linkers) accommodates two electrons. The formal electronic configuration of the Fe atom is thus [Ar] $3d^6$, with the d-orbital occupancy of $(d_{xz})^1 (d_{yz})^1 (d_{xy})^1 (d_{x2-y2})^1 (d_{z2})^2$, as summarized in Table 1A. As evident from the DOS plots and Table 1, five electrons reside in the majority spin channel, while only one electron in the $(d_{z2})$ orbital is in the minority spin channel; the spin of the $Fe^{2+}$ atom is thus S=2 (quintet).

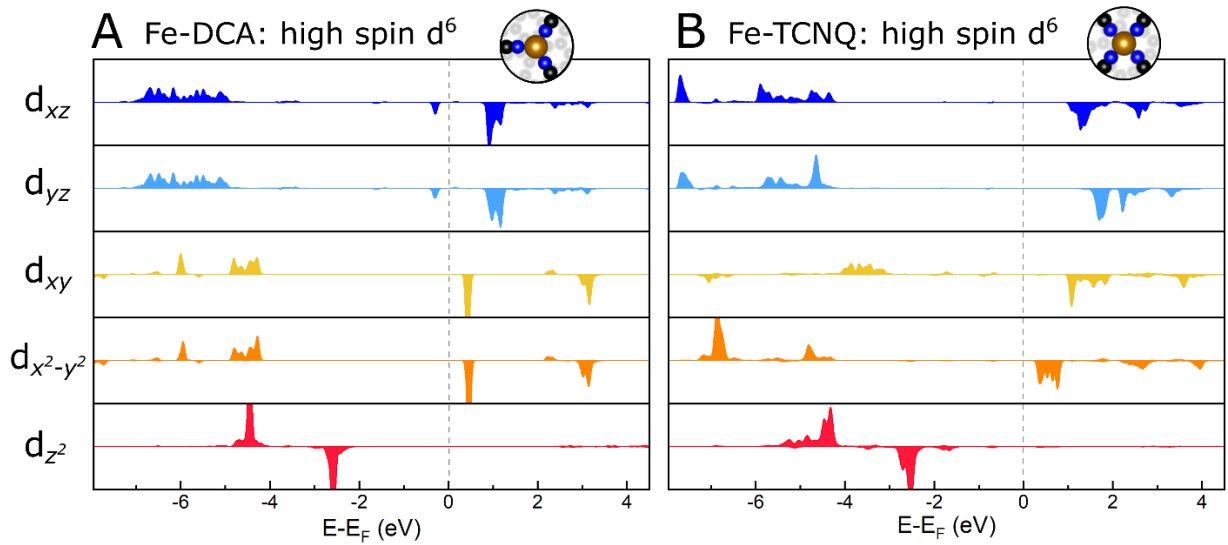

Figure 2: Density of states (DOS) projected on the Fe 3d orbitals of the Fe-$N_3$ and Fe-$N_4$ sites within Fe-DCA (A) and Fe-TCNQ (B). In both systems, the electronic configuration is high-spin $d^6$, with the $d_{xz}$, $d_{yz}$, $d_{xy}$ and $d_{x2-y2}$ orbitals being singly-occupied, and the $d_{z2}$ being doubly-occupied.

Table 1: Fe 3d orbital occupancies (A) and centers of mass (B) calculated for the Fe-DCA and Fe-TCNQ systems supported on graphene.

| A | Fe d-orbital occupancy ($e^-$) | | | | | |
|---|---|---|---|---|---|---|
| | All d-orbitals | $d_{xz}$ | $d_{yz}$ | $d_{xy}$ | $d_{x2-x2}$ | $d_{z2}$ |
| Fe-DCA/gr | 6.09 | 1.05 | 1.04 | 1.07 | 1.08 | 1.86 |
| Fe-TCNQ/gr | 6.05 | 1.00 | 1.00 | 1.19 | 1.00 | 1.85 |

| B | Fe d-orbital center of mass (eV) | | | | | |
|---|---|---|---|---|---|---|
| | All d-orbitals | $d_{xz}$ | $d_{yz}$ | $d_{xy}$ | $d_{x2-x2}$ | $d_{z2}$ |
| Fe-DCA/gr | -2.78 | -2.24 | -2.22 | -2.91 | -2.93 | -3.55 |
| Fe-TCNQ/gr | -2.63 | -2.21 | -2.03 | -2.79 | -2.48 | -3.61 |

The same electronic configuration is found by DFT computations of Fe-$N_4$ sites within Fe-TCNQ (Figure 2B, and Table 1A), and this result is also consistent with recent XAS study.[49] In the quasi-tetrahedral Fe-$N_4$ site the hybridization between the Fe d-orbitals and TCNQ states involves mainly the $d_{xz,yz}$ orbitals which have the highest spatial overlap with the N p orbitals. Same as in the case of Fe-DCA, all the d orbitals save for $(d_{z2})^2$ are singly-occupied, leading to the formal electronic configuration of [Ar] $3d^6$, with the d-orbital occupancy of $(d_{xz})^1 (d_{yz})^1 (d_{xy})^1 (d_{x2-y2})^1 (d_{z2})^2$ and spin S=2.

The electronic configuration of the Fe-$N_3$ and Fe-$N_4$ sites within Fe-DCA and Fe-TCNQ is almost identical, but it is well known that the reactivity of transition metals depends not only on the filling of the d-orbitals, but also on their position with respect to Fermi level.[21,33,34,50] This parameter is commonly quantified by the center of mass of the whole d-band or of the individual d-orbitals, and the dependence of the d-band position on reactivity is commonly referred to as the "d-band center model". In Table 1B, we summarize the centers of the d-band as well as the centers of individual d-orbitals for both systems. We find that the positions of the d-orbital centers are very similar, albeit slightly higher in Fe-TCNQ. This applies to the d-orbitals expected to be involved in bonding with upright-standing adsorbates ($d_{z^2}$, $d_{xz}$, $d_{yz}$) as well as the ones expected to be non-bonding ($d_{xy}$, $d_{x^2-y^2}$) Thus, based on the electronic structure parameters, one is tempted to assume that the Fe atoms within these systems should bind reactants like carbon monoxide (CO) with similar strength, perhaps slightly more strongly in the Fe-N$_4$ coordination. However, the opposite is true, as shown both by theory and experiment below.

### *CO adsorption on Fe-N$_3$ and Fe-N$_4$ models*

Carbon monoxide is a crucially important molecule for both industrial applications and fundamental studies. Industrial applications utilize CO as a reactant in preferential CO oxidation (PROX),[51] water-gas shift[52] or Fischer-Tropsch synthesis,[53] while in fundamental studies CO is widely used as a probe molecule, whose C-O stretching frequency is used to evaluate charge states, coordination numbers and chemical reactivities of model SACs.[14,37,54,55]

Our DFT data presented in Figure 3 show that CO binds to Fe-N$_3$ and Fe-N$_4$ sites in a qualitatively similar way, but the CO adsorption energies ($E_{ads\_CO}$) significantly differ, being −0.84 eV on Fe-N$_3$, but only −0.23 eV on Fe-N$_4$. The mechanism of Fe-CO bonding on Fe-N$_x$ is consistent with the picture known from studies of extended metal surfaces:[34,50,56,57] The σ-bond is facilitated by highest-occupied molecular orbital of CO (5σ) that hybridizes with the $d_{z^2}$ orbital of the Fe atom, while the π-bond back-donation takes place through hybridization of the lowest unoccupied, doubly degenerate molecular orbital of CO (2π*) with the $d_{xz}$, $d_{yz}$ orbitals of Fe. Both these effects are clearly observed in DOS plots shown in Figure 3. While the σ-bonding appears very similar in both cases, there is a significant difference in the π-bonding, which is much more pronounced in the Fe-DCA case (2π*/$d_{xz/yz}$ peaks at -2.9 eV and -1.1 eV) than in the Fe-TCNQ case (2π*/$d_{xz/yz}$ peaks at -2.6 eV and -0.8 eV). This qualitative observation is supported by the computed occupancy of the 2π* orbital, which is 0.72 e$^-$ in the CO/Fe-DCA case, but only 0.36 e$^-$ in CO/Fe-TCNQ. The higher occupancy of the 2π* orbital also results in increased C-O bond length;[34,56] which is 116 pm on CO/Fe-DCA and 114 pm on CO/Fe-TCNQ. All this indicates that the covalent bond between CO and Fe is indeed significantly stronger in the CO/Fe-DCA case than in CO/Fe-TCNQ, and this conclusion is further supported by additional analyses presented in Figures S3-S7 in the SI. The observation of a stronger Fe-CO bond in CO/Fe-DCA is in line with the computed $E_{ads\_CO}$ values, but in disagreement with the expectations based on d-band center positions presented in Table 1B.

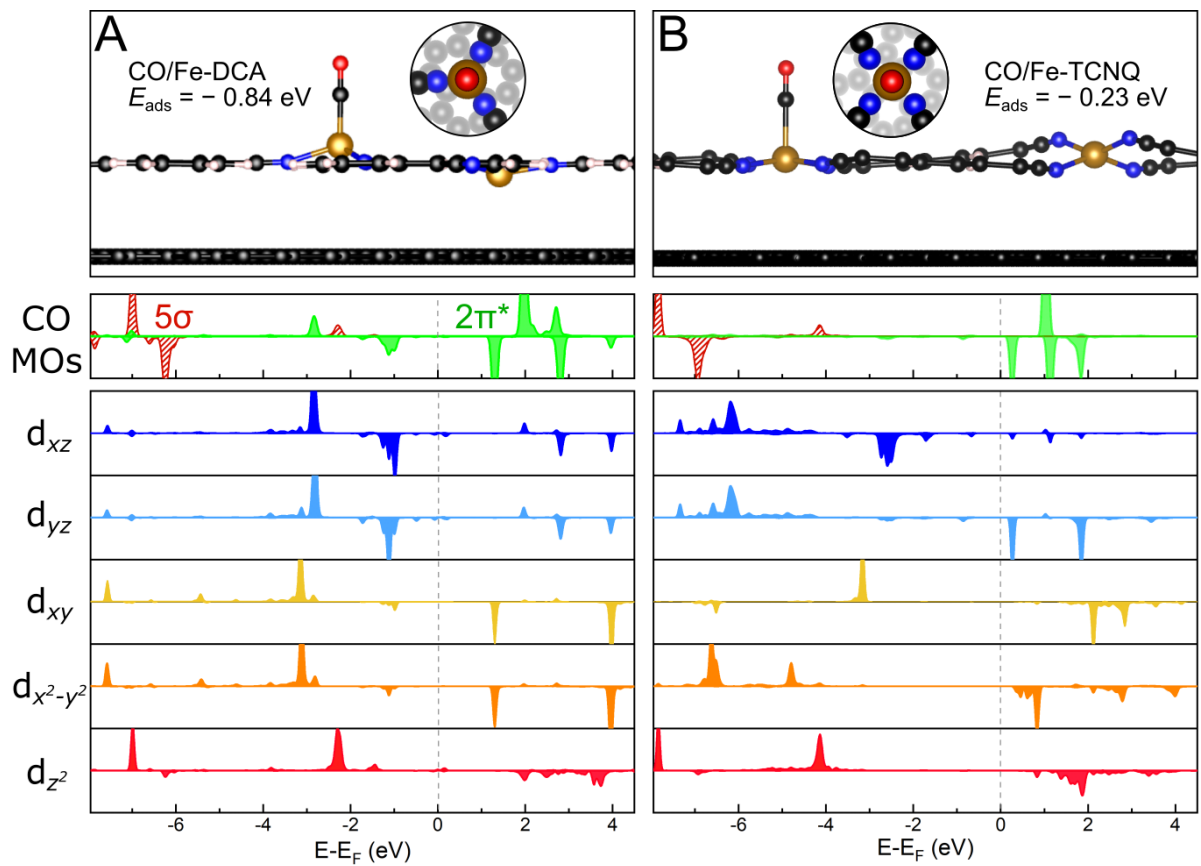

Figure 3: Computed physical and electronic structures of CO/Fe-DCA and CO/Fe-TCNQ systems supported in graphene. (A) Upon CO adsorption the Fe cation is significantly lifted from its initial Fe-$N_3$ coordination. The binding energy is found to be −0.84 eV. The density of states plot indicates that the covalent Fe-CO bonding occurs through hybridization of 5σ CO orbital with Fe $3d_{z^2}$, and 2π* CO orbital with Fe $3d_{xz/yz}$. (B) The Fe-$N_4$ site is initially quasi-tetrahedral, but upon CO adsorption it gets planarized. The Fe-CO bonding is qualitatively similar to the Fe-$N_3$ case, but the adsorption is significantly weaker (−0.23 eV). The Fe $3d_{xz/y}$-CO 2π* bonding is significantly less pronouced, as seen in the much lower occupancy of the CO 2π* orbital.

We carried out variable-temperature STM measurements to provide experimental confirmation of the DFT results. Figure 4A shows an STM image of Fe-DCA/graphene acquired at 128 K prior to CO exposure, panels B,C show STM images acquired on the same area in a background CO pressure of 2×10$^{-8}$ mbar. In subsequent STM images acquired in a CO background, the adsorbed CO molecules are imaged as bright features appearing one-by-one on the Fe-$N_3$ sites.

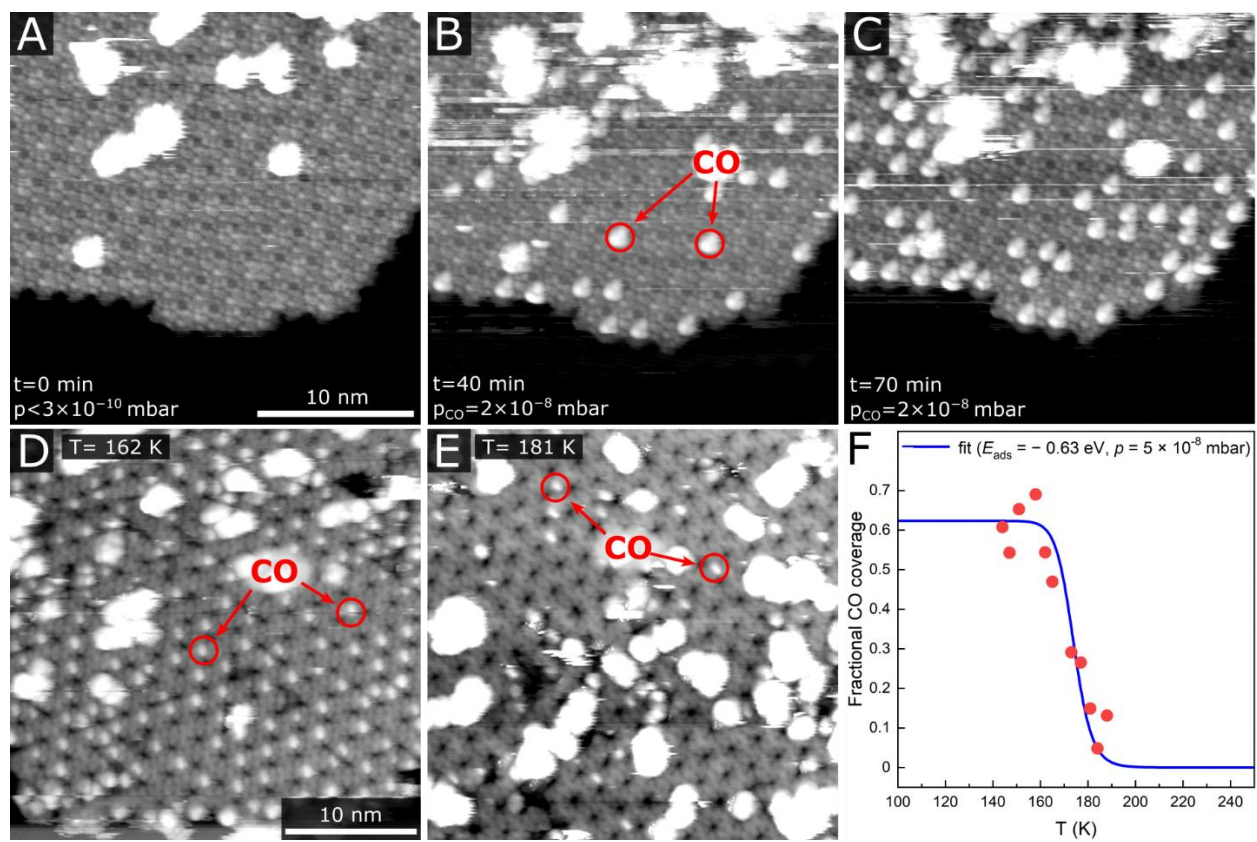

Figure 4: Following adsorption and desorption of CO on Fe-N$_3$ sites within Fe-DCA by variable-temperature STM. (A) STM image of the clean Fe-DCA/gr acquired at 128 K prior to CO dosing. ($V_{bias}$= −1.0 V, $I_{tunnel}$= 50 pA). (B) STM image taken at 131 K, after ≈40 min exposure to 2×10$^{-8}$ mbar CO (about 35 Langmuir dose). The CO molecules are observed as bright features appearing on the Fe-N$_3$ sites ($V_{bias}$= −0.9 V, $I_{tunnel}$= 50 pA). (C) STM image taken at 133 K, after ≈70 min of CO exposure (about 60 Langmuir dose). The number of CO molecules is increasing with higher CO exposure ($V_{bias}$= −0.7 V, $I_{tunnel}$= 50 pA). (D-F) Variable-temperature STM allows to follow the change in equilibrium coverage of CO on Fe-N$_3$ sites. (D) STM image acquired in a CO background ($p_{CO}$=5×10$^{-8}$ mbar) at 162 K. Many Fe-N$_3$ sites are capped with a CO molecule. (E) STM image acquired in the same CO background pressure at a temperature of 181 K. The number of adsorbed CO molecules is significantly reduced. (F) Analysis of temperature-dependent CO coverage on the Fe-DCA/gr, measured in a CO background of 5×10$^{-8}$ mbar. The red dots are experimental data extracted from STM images, the blue curve is a Langmuir isobar fit, calculated for $E_{ads}$=−0.63 eV and background pressure of 5×10$^{-8}$ mbar and normalized for maximum observed coverage.

Having identified the CO/Fe-N$_3$ species, we increased the CO background pressure to 5×10$^{-8}$ mbar, and we continuously imaged the Fe-DCA 2D MOF while heating the sample with a slow temperature ramp of 0.2-0.3 K/min. Figure 4D shows an image taken at 162 K, where a high coverage of CO molecules on Fe-N$_3$ sites is observed. As the temperature rises, the fraction of occupied Fe sites decreases, as seen in Figure 4E showing an STM image taken at 181 K. We analyzed the fractional coverage of CO on Fe-N$_3$ sites in subsequent STM images taken at temperatures between 144 – 188 K, and we used equilibrium thermodynamics approach to calculate the CO adsorption energy from this experimental dataset. Figure 4C shows experimental data fitted with a Langmuir isobar calculated for the experimental CO pressure of 5×10$^{-8}$ mbar. The $E_{ads\_CO}$ was treated as a variable parameter of the fit (see SI for details), and the best agreement was found for $E_{ads\_CO} = -0.63$ eV. This value is consistent with the computational prediction, considering the well-known fact that the DFT methodology used typically overestimates the CO binding by about 0.2 eV[21,34,50] due to underestimation of the CO HOMO-LUMO gap.[58]

The same set of experiments was carried out on the Fe-TCNQ/graphene system, but no evidence of CO adsorption was observed at temperatures down to 115 K. Figure 6A shows that the Fe-$N_4$ sites remain free of CO even when scanned in a 5×10$^{-8}$ mbar CO background at 118 K. The image shows three rotational domains of Fe-TCNQ monolayer and an island of Fe-TCNQ bilayer in a top left corner.[59] The monolayer Fe-TCNQ domains feature a characteristic "zig-zag" pattern, indicating that the Fe-$N_4$ sites remain in a CO-free quasi-tetrahedral coordination (see ref [47] for details). This agrees with computational predictions: Figure 5B shows calculated Langmuir isobar for $E_{ads\_CO}$ values of −0.2 eV and −0.3 eV, indicating that the Fe-$N_4$ sites are only expected to stably coordinate CO molecules at temperatures well below 100 K in the CO pressure of 5×10$^{-8}$ mbar. The experimental observations are thus fully consistent with computations, unambiguously proving that the CO adsorption energy is significantly higher on Fe-$N_3$ sites within Fe-DCA than on the Fe-$N_4$ sites within Fe-TCNQ.

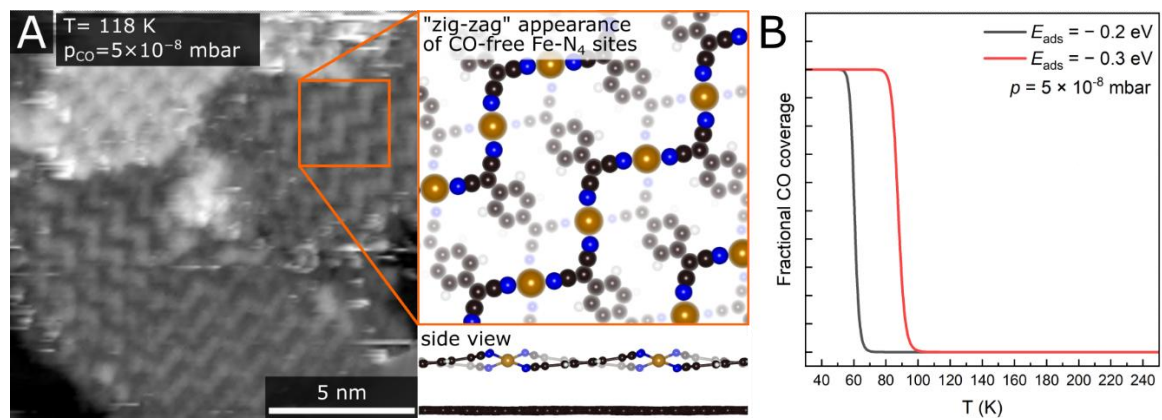

Figure 5: Carbon monoxide does not adsorb on the Fe-$N_4$ sites within Fe-TCNQ at 118 K. (A) Scanning tunneling microscopy image of Fe-TCNQ, acquired at 118 K in a background CO pressure of 5×10$^{-8}$ mbar (total CO dose higher than 100 Langmuir). The characteristic "zig-zag" appearance is observed on the monolayer Fe-TCNQ areas, which clearly indicates that the Fe-$N_4$ sites remain in the quasi-tetrahedral coordination, and no CO is adsorbed. (B) Langmuir isobar calculated for the experimental CO pressure and $E_{ads\_CO}$ of −0.3 eV and −0.2 eV. The CO molecules are only expected to be stably adsorbed at temperatures below 100 K (resp. below 70 K for $E_{ads\_CO}$=−0.2 eV).

### The origin of different reactivity of Fe-$N_3$ and Fe-$N_4$ sites

The observed difference in Fe-$N_3$ and Fe-$N_4$ reactivity does not originate from the ground state electronic structures of the Fe-$N_x$ sites, because these are almost identical, as shown in Figure 2 and Table 1. Therefore, in the following section we focus on the effect of the CO-induced structural relaxations that are present in both models. Specifically, in the Fe-DCA case the CO ligation lifts the Fe atom up from the -$N_3$ plane to form a tetrahedral geometry, while the quasi-tetrahedral Fe-$N_4$ sites within Fe-TCNQ get planarized upon CO adsorption. The computed

$E_{ads\_CO}$ values contain contributions from both the Fe-CO bond ($E_{bond}$) and from the structural distortion ($E_{MOF\_distortion}$):

$$E_{ads\_CO} = E_{bond} + E_{MOF\_distortion}$$

where $E_{bond}$ can be calculated as the energy difference between the most stable structure with CO and the same structure frozen in the CO-induced geometry with the CO molecule removed to the gas phase. The $E_{MOF\_distortion}$ is calculated as the energy difference between the most stable MOF structure and the structure frozen in the distorted geometry after removing the CO. The values of $E_{ads\_CO}$, $E_{bond}$ and $E_{MOF\_distortion}$ are summarized in Table 2. The computations were carried out both in a free-standing approximation and supported on graphene, this allows to rigorously quantify any effects induced by the interaction with the support. The data show that CO-induced lifting of the Fe atom within the Fe-DCA/graphene structure costs 0.18 eV, while the Fe-CO bond stabilizes the system by −1.02 eV. Together, these contributions add up to $E_{ads\_CO}$ of −0.84 eV. The value of $E_{ads\_CO}$ is very similar between an ideal free-standing and graphene-supported Fe-DCA models, but in the free-standing approximation the lack of van der Waals interaction with the support allows larger CO-induced structural distortions; this results in different values of $E_{bond}$ and $E_{MOF\_distortion}$.

Table 2: Computed CO adsorption energies ($E_{ads\_CO}$), energies of the Fe-CO bond ($E_{bond}$) and energies required to distort the MOF structures to accommodate the CO adsorbate ($E_{MOF\_distortion}$) on the individual Fe-DCA and Fe-TCNQ models

|  | $E_{ads\_CO}$ | $E_{bond}$ | $E_{MOF\_distortion}$ |
|---|---|---|---|
| free-standing Fe-DCA | − 0.76 eV | −1.22 eV | 0.46 eV |
| Fe-DCA/graphene | −0.84 eV | −1.02 eV | 0.18 eV |
| free-standing Fe-TCNQ | −0.20 eV | −0.41 eV | 0.21 eV |
| non-planar Fe-TCNQ/graphene | −0.23 eV | −0.42 eV | 0.19 eV |
| planar Fe-TCNQ/graphene | −0.32 eV | −0.41 eV | 0.09 eV |

The CO-induced structural relaxation is qualitatively the opposite in Fe-TCNQ, as the CO-free structure features Fe in non-planar Fe-N$_4$ sites, and the CO ligation causes their planarization. Therefore, the energy required for the Fe-N$_4$ planarization depends on the initial structure. When the $E_{MOF\_distortion}$ is referenced to a free-standing structure with Fe in perfect tetrahedrons, the energy required to planarize one Fe site is 0.21 eV, and this value is very similar (0.19 eV) when a slightly planarized Fe-TCNQ/graphene structure is taken as a reference (shown in Figure 1E). Finally, when a hypothetical perfectly planar structure is assumed, the structural changes induced by CO adsorption cost only 0.09 eV. Regardless of the different energy references (which lead to different $E_{ads\_CO}$ values from −0.20 eV to −0.32 eV), the strength of the Fe-CO bond remains remarkably consistent in these different models, falling within (−0.410 ± 0.006) eV. This is because the coordination geometry of the final CO/Fe-N$_4$ structure is the same in all the tested Fe-TCNQ models.

To summarize, it is found that the cost of structural relaxations ($E_{\text{MOF\_distortion}}$) is comparable across the different graphene-supported Fe-N$_3$ and Fe-N$_4$ models. This indicates that the observed differences in $E_{\text{ads\_CO}}$ do *not* stem from different electronic structures of the Fe cations, *nor* from the different energetic costs of adsorbate-induced relaxations. However, the structural distortions do not only cost energy; they can simultaneously stabilize the Fe-CO bonding by allowing more efficient hybridization. This effect is clearly seen in Figure 6A, where we plot the $E_{\text{ads\_CO}}$, $E_{\text{bond}}$ and $E_{\text{MOF\_distortion}}$ values of CO/Fe-DCA models, whose structures were kept fixed in intermediate points between the CO-free and CO-adsorbed structures. When the Fe-DCA structure is fixed to be planar ($\Delta z = 0.0$ Å), then the $E_{\text{bond}}$ is −0.41 eV, i.e. identical to the $E_{\text{bond}}$ value found in Fe-TCNQ models. However, once the Fe atom is lifted above the plane, the $E_{\text{bond}}$ is significantly strengthened, by about (−0.12 ± 0.03) eV per every 0.1 Å increase of $\Delta z$. The price paid for the Fe-CO bond stabilization is the $E_{\text{MOF\_distortion}}$, but this is very low for small $\Delta z$, and only starts to significantly increase at $\Delta z \geq 0.3$ Å. The minimum energy of the system is found at $\Delta z = 0.6$ Å, which is close to the structural distortion of the fully relaxed system ($\Delta z_{\text{rel}} = 0.68$ Å). At this point the total $E_{\text{ads\_CO}} = -0.81$ eV is very close to the value found in fully-relaxed structures (−0.84 eV).

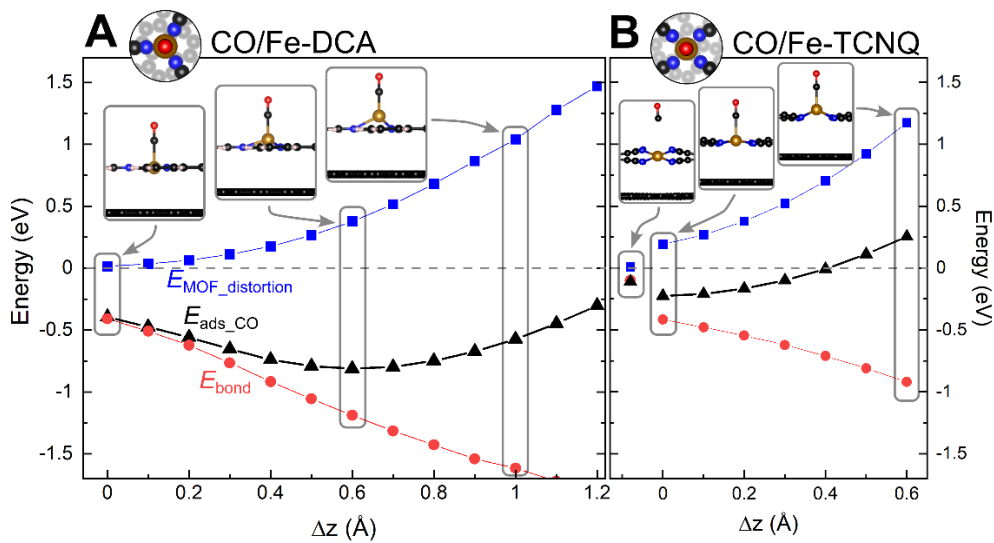

Figure 6: Energetics of the Fe-CO bonding as a function of the Fe-lifting above the -N$_x$ plane ($\Delta z$) for the CO/Fe-DCA (A) and CO/Fe-TCNQ (B) models. When the structures are fixed in a planar configuration ($\Delta z=0.0$Å), the $E_{\text{bond}}$ values of the two models are identical. Lifting the Fe atom above the -N$_x$ plane significantly stabilizes the $E_{\text{bond}}$ in both systems. The main difference lies in the $E_{\text{MOF\_distortion}}$, which rises much slower in Fe-DCA than in Fe-TCNQ. The minimum of the total CO adsorption energy ($E_{\text{ads\_CO}}$) thus lies at $\Delta z=0.6$Å for Fe-DCA, where the $E_{\text{bond}}$ is significantly stronger (−1.19 eV). In contrast, the quickly rising $E_{\text{MOF\_distortion}}$ prevents such structural change in Fe-TCNQ, and the minimum of $E_{\text{ads\_CO}}$ lies at $\Delta z=0.0$Å.

A similar analysis was carried out on the CO/Fe-TCNQ system to test if a similar Fe-CO bond stabilization mechanism could take place here. First, we tested the ground-state Fe-TCNQ structure with the Fe-N$_4$ sites being fixed in the quasi-tetrahedral geometry. The $E_{ads\_CO}$ value of only −0.10 eV highlights again the crucial role of structural flexibility: without the possibility to planarize the Fe-N$_4$ structure, the site is inert to any chemical interaction with CO. As the next step, we have taken the planarized CO-induced geometry as a reference structure ($\Delta z = 0.0$ Å), and we lifted the Fe atom from this reference. As shown in Figure 6B, this stabilizes the Fe-CO bond in a similar way as in the Fe-N$_3$ site, by about (−0.08 ± 0.02) eV per every 0.1 Å increase of $\Delta z$. However, the $E_{MOF\_distortion}$ rises faster already from low $\Delta z$, making this Fe-CO bond stabilization mechanism unfavorable. This presents the fundamental difference between the two sites that results in significantly stronger CO adsorption on the Fe-N$_3$ compared to Fe-N$_4$ sites.

*Discussion*

Overall, our results reveal that the fundamental origin of the distinct reactivities of Fe-N$_3$ and Fe-N$_4$ lies in their different structural flexibility. When the Fe-N$_3$ and Fe-N$_4$ sites are fixed in similar (planar) geometries, then the CO binding energies are identical (−0.410 ± 0.006 eV), as one would expect based on their identical electronic configuration, similar d-orbital center positions and minimal steric hindrance for CO adsorption. However, this picture dramatically changes once the structures are allowed to relax. The structural flexibility of the Fe-N$_3$ site allows lifting of the Fe cation by 0.68 Å, which costs 0.18 eV but strengthens the Fe-CO bond by additional 0.61 eV. In contrast, on the Fe-N$_4$ site a price of 0.19 eV is paid to planarize the Fe-N4 geometry, and while an additional Fe-CO bond stabilization is possible by lifting of the Fe atom from the -N$_4$ plane, it is energetically unfavorable due to the structural rigidity of the Fe-N$_4$ geometry.

Having identified the physical origin of the differing Fe-CO bond strengths, we further analyzed how the observed structural changes correlate to the electronic structure parameters of the CO-free models. In both models, lifting of the Fe atom enables stabilization of both the Fe 3d$_{z^2}$ - CO 5σ and Fe 3d$_{xz/yz}$ - CO 2π* bonds, but the effect is significantly stronger on the π bonding (detailed in Figures S3-S7 in the SI). Thus, within the d-band center model framework one would expect that such stronger Fe-CO bonding could be correlated to an up-shift of the Fe d$_{xz/dyz}$ and d$_{z^2}$ centers in CO-free structures. However, our data do not indicate such a scenario; the Fe d-orbital center positions of CO-free structures do not follow a consistent trend upon Fe-lifting (detailed in Figure S8 in the SI). This means that the observed trend of increasing Fe-CO bond strength upon Fe-lifting is not easily correlated to the electronic parameters of CO-free structures in their electronic ground states.

All this implies that the reactivity descriptors based on electronic parameters of the ground state structures have limited applicability for SAC systems. The different local environments of the SAC sites inevitably lead to different possibilities for structural relaxation, and such relaxation can significantly affect the adsorption energetics. The coordination number of the metal and the nature of the metal-ligand bonds are the most obvious parameters defining the local environment, but significant differences should also be expected between sites with the same coordination number and same ligands: for example, in TCNQ-based 2D MOFs the metal sites are typically coordinated to four cyano- groups, but the preference for square-planar or tetrahedral local geometry varies from cation to cation. While the Ni cations in Ni-TCNQ prefer geometries close to square-planar,[60] a similar Co-TCNQ system shows preference for tetrahedral Co sites.[61] Such differences will significantly affect both the energy required for planarization and the energy required for potential cation lifting. Structural flexibility can thus significantly vary even in very similar systems, implying the need for analyzing the individual systems on a case-by-case basis, and establishing general trends from reliable quantitative datasets, where both the electronic properties and structural flexibility are considered. This makes computational screening for promising SACs a much more complex task, but it significantly increases the chances of finding efficient SACs.

## *Conclusions*

This work quantifies the reactivity differences between model single-atom catalyst sites with identical electronic structure (high-spin $d^6$) but different local coordination geometry. It is found that threefold coordinated Fe-$N_3$ sites are significantly more reactive towards carbon monoxide than fourfold coordinated Fe-$N_4$ sites; the difference in adsorption energy is 0.62 eV. This difference can be tracked down to distinct possibilities of structural relaxations: the Fe cation in trigonal-planar geometry can be easily lifted above the plane, which costs 0.18 eV but strengthens the Fe-CO bond by 0.61 eV, primarily through increased Fe $3d_{xz/yz}$ - CO $2\pi^*$ back-bonding. In contrast, a similar mechanism is unfavorable in Fe-$N_4$ sites due to the structural rigidity of the square-planar geometry. These results demonstrate the crucial role of coordination geometry in single atom catalysis, and that these effects are not captured by reactivity descriptors based solely on the electronic structure parameters.

## *Acknowledgements*

Z.J. acknowledges funding from the Czech Science Foundation (GAČR grant 24-10593L). J.P. was supported by the European Union's Horizon Widera 2022 program under grant agreement No. 101130765. The DFT computations were carried out at IT4I National Supercomputing Center (supported by MEYS CR through the e-INFRA CZ, ID: 90140). The CzechNanoLab project LM2023051, funded by MEYS CR, is acknowledged for the financial support of the research at CEITEC Nano Research Infrastructure.

*Contributions*

Z.J., D.H., T.L., A.J. acquired and analyzed the experimental data, J.P. and T.L. carried out the computational analysis. Z.J. and J.Č. conceptualized and supervised the research. Z.J. and J.Č. acquired the research funding. Z.J. and J.P. wrote the manuscript with the input from all co-authors.

*Supporting information*

Experimental and computational methods, additional STM data and additional computational analysis of the Fe-CO bonding in the Fe-$N_3$ and Fe-$N_4$ models. The supporting information file contains additional references.[62–77]

**Supplementary Information**

# Structural flexibility dictates reactivity of single-atom catalysts

Jakub Planer[1], Dominik Hrůza[1], Tadeáš Lesovský[1], Ayesha Jabeen[1], Jan Čechal[1,2*], Zdeněk Jakub[1*]

[1] CEITEC – Central European Institute of Technology, Brno University of Technology, Purkyňova 123, 61200 Brno, Czechia

[2] Faculty of Mechanical Engineering, Brno University of Technology, Technická 2896/2, 616 69 Brno, Czechia

Correspondence to: cechal@fme.vutbr.cz, zdenek.jakub@ceitec.vutbr.cz

# Table of Contents

# Experimental Methods

Experiments were carried out in an ultrahigh vacuum system consisting of multiple chambers interconnected by a central transfer line, separated by gate valves. The base pressure of all the chambers used in this study is below $5 \times 10^{-10}$ mbar. The Ir(111) single crystals were purchased from MaTecK and SPL, the Ir(111) surface was prepared by cycles of Ar$^+$ sputtering (1.5 keV, 10 min) and annealing up to 1300 °C for 10 min. When graphene was present on the sample prior to cleaning, the first annealing cycle took place in O$_2$ background (up to 1100 °C, $p_{O2} = 1 \times 10^{-6}$ mbar). The temperature was measured by a LumaSense IMPAC IGA 140 pyrometer with the emissivity set to 0.1.

Graphene on Ir(111) was grown by adsorbing saturation coverage of ethylene at room temperature, followed by pumping out the ethylene background and ramping the temperature up to 1250 °C in UHV. At 1250 °C the sample was re-exposed to ethylene ($5 \times 10^{-7}$ mbar, 5 min). This protocol combines temperature-programmed growth (TPG) with chemical vapor deposition (CVD) and consistently leads to a full monolayer coverage of high-quality graphene/Ir(111).[1,2]

TCNQ was evaporated from a quartz crucible heated to 115 °C (MBE Komponenten OEZ), Fe was evaporated from an alumina crucible heated to 1030 °C (MBE Komponenten HTEZ). The evaporation rate of Fe was checked by a water-cooled quartz crystal microbalance. The sample temperature during Fe-TCNQ synthesis was calibrated by a special sample holder with a K-Type thermocouple attached close to the crystal surface. The Fe-TCNQ synthesis protocol involved saturation by TCNQ at a ≈ 80 °C, followed by co-deposition of Fe and TCNQ at the same temperature. After the co-deposition step, the samples were annealed to 420 °C. This annealing step leads to improved long-range order of the Fe-TCNQ 2D MOF and also leads to the desorption of TCNQ molecules adsorbed atop the 2D MOF.[3] A detailed analysis of graphene-supported Fe-TCNQ systems is provided in refs [4,5].

DCA was evaporated from a quartz crucible placed in a Createc NATC evaporator heated by a heat transfer fluid kept at 110 °C (using a Huber CC304 bath thermostat). The sample temperature during Fe+DCA codeposition was ≈ 70 °C, afterwards the sample was post-annealed up to ≈ 90 °C.

Scanning tunneling microscopy images were recorded in the constant current mode using a variable-temperature SPECS Aarhus 150 system equipped with a tungsten tip. During low-temperature operation, the sample stage is cooled by liquid nitrogen flow through a stage-locking piston, the thermal contact during STM scanning is facilitated by copper braids. The temperature of the sample rises during STM scanning, which results in large thermal drift. The image distortion was corrected during post-processing to fit the known dimensions of the measured structures: the unit cells of the gr/Ir moiré, Fe-TCNQ and Fe-DCA 2D MOFs. Where possible, nonlinear image distortion was corrected as described in ref. [6]. Carbon monoxide was dosed directly to the STM chamber by a high-precision leak valve, the pressure was monitored by a Bayard-Alpert gauge.

Low Energy Electron Diffraction data were acquired in a SPECS FE-LEEM P90 instrument, diffraction patterns were modelled and visualized using ProLEED Studio software.[7,8]

**Langmuir isobar analysis**

The Langmuir isobar shown in Figures 4C and 5B in the main text is given by the expression

$$\theta(T) = \frac{1}{\exp\left(\frac{E_{ads\_CO} - \mu(T)}{kT}\right) + 1},$$

where $\theta(T)$ is the normalized fractional CO coverage, $E_{ads\_CO}$ is the CO adsorption energy, $k$ is Boltzmann constant, $T$ is temperature and $\mu(T)$ is the equilibrium chemical potential of the gas-phase CO and adsorbed CO which can be expressed as

$$\mu(T) = kT\left[\ln\frac{ph^3}{(2\pi mkT)^{3/2}kT} + \ln\left(\frac{T}{T_R}\right)\right],$$

where $p$ is the pressure of CO, $h$ is Planck constant, $m$ is the mass of the CO molecule and $T_R = 2.78$ K is the rotational temperature of CO. In the analysis presented in Figure 4C, the isobar was evaluated for the experimental CO pressure of $5\times10^{-8}$ mbar, while the $E_{ads\_CO}$ was treated as a variable parameter of the fit.

# Computational Methods

Spin-polarized density functional theory (DFT) calculations were performed with the Vienna *ab initio* Simulation Package (VASP)[9] using the projector augmented wave method (PAW)[10] to treat core electrons. We used a nonlocal van der Waals corrected optPBE-vdW functional[11] for the description of exchange correlation energy. A Hubbard-like coulomb repulsion correction $U$-$J$ = 4 eV in Dudarev's formulation[12] was employed for an appropriate description of Fe 3d orbitals. For iron, nitrogen, oxygen, hydrogen and carbon, 11 valence electrons ($3s^23p^64s^23d^6$), 5 valence electrons ($2s^22p^3$), 6 valence electrons ($1s^22p^4$), 1 valence electron ($1s^1$), and 4 valence electrons ($2s^22p^2$), respectively, were expanded in a plane-wave basis set with an energy cutoff set to 520 eV. The Brillouin zone was sampled with a 3×3×1 Γ-centered Monkhorst–Pack grid.[13] Geometry optimization was stopped when all residual forces acting on atoms were less than 0.02 eV/Å.

We used the Occupation Matrix Control[14] (OMC) of the Fe 3d manifold in the Fe-DCA networks to generate the initial electronic configuration with correct occupation matrices (corresponding to the electronic ground state). For the three-fold coordinated Fe atom, the occupation matrix in the spin-up channel was set close to the identity matrix, while for the spin-down channel, one electron was assigned to the $d_{z^2}$ orbital (see eq. 1). This electronic configuration was subsequently relaxed without the OMC. This procedure allowed us to reliably obtain the most stable electronic configuration. The occupation matrices for the spin-up and for the spin-down channel employed in calculations are:

$$M_\uparrow = \begin{pmatrix} 0.98 & 0 & 0 & 0 & 0 \\ 0 & 0.95 & 0 & 0 & 0 \\ 0 & 0 & 0.95 & 0 & 0 \\ 0 & 0 & 0 & 0.95 & 0 \\ 0 & 0 & 0 & 0 & 0.98 \end{pmatrix}, M_\downarrow = \begin{pmatrix} 0.1 & 0 & 0 & 0 & 0 \\ 0 & 0.1 & 0 & 0 & 0 \\ 0 & 0 & 0.92 & 0 & 0 \\ 0 & 0 & 0 & 0.1 & 0 \\ 0 & 0 & 0 & 0 & 0.1 \end{pmatrix} \quad (1)$$

For interface calculations, a single graphene sheet was used as a substrate. To create the interface models, we kept the substrate atoms at their positions calculated from DFT, adjusted the molecular layer accordingly, and added a 15 Å thick vacuum layer. For the Fe-TCNQ/gr interface, we proceed from the supercell employed in our previous work,[5] but doubled in the $b$ direction in order to avoid interactions between the adsorbates over the periodically repeated replicas. The matrix notation of the employed supercell is $\begin{pmatrix} -4 & -13 \\ 10 & 2 \end{pmatrix}$ for graphene and $\begin{pmatrix} 4 & 0 \\ 0 & 2 \end{pmatrix}$ for the Fe-TCNQ layer, resulting in Fe-TCNQ lattice parameters that lie between those of the gas-phase tilted configuration and the planarized configuration.[5]

The Fe-DCA/gr interface was modelled using the supercell with the matrix notation $\begin{pmatrix} 7 & -2 \\ 9 & 2 \end{pmatrix}$ for graphene, which was merged with the primitive unit cell of the honeycomb-kagomé Fe-DCA. The dimensions of the supercell are 0.8% larger than the lattice parameter of the FeDCA gas phase model.

The Crystal Orbital Hamilton Population (COHP) analysis of the Fe-CO bonding was calculated using the LOBSTER code,[15,16] and the LUMO occupations were calculated as integrated projections of a full system wavefunction onto a gas-phase CO Kohn-Sham orbitals, following the methodology described in reference [17].

## Additional STM data of Fe-DCA/gr

Figure S1 shows a large-scale STM image of Fe-DCA/gr. The Fe-DCA structure is observed in two distinct rotations with respect to the graphene/Ir moiré, in agreement with the LEED data presented in the main text. The characteristic dimensions of the Fe-DCA islands are higher tens of nm. The bright features on top of the Fe-DCA islands are most likely excess Fe+DCA species; further annealing up to 110 °C did not lead to desorption of these features.

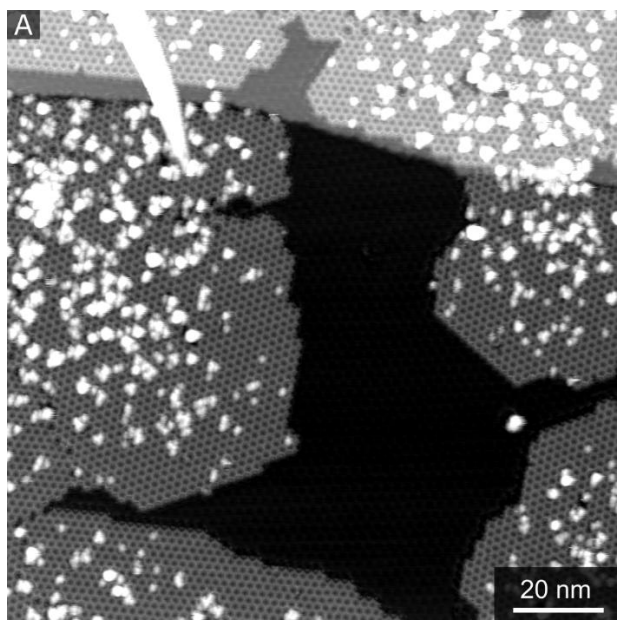

Figure S1: Large-scale STM image of the Fe-DCA/gr/Ir(111) system (135 × 135 nm, $U_{sample} = -1.8$ V, $I_{tunnel} = 0.05$ nA)

## Additional STM data of Fe-TCNQ/gr

The STM appearance of the Fe-TCNQ/gr/Ir(111) system strongly depends on the applied bias voltage, as described in detail in reference [5]. At sample bias around – 0.6 V (Figure S2A), the Fe-TCNQ monolayer looks homogeneous, but at sample bias close to – 1.3 V (Figure S2B), characteristic "zig-zag" appearance is observed on parts of the monolayer Fe-TCNQ areas. This is because at these tunneling conditions the most intense signal originates from the electronic states located at the cyano- and methyl- groups of the TCNQ molecule, which are physically corrugated within the quasi-tetrahedral Fe-N$_4$ sites. The quasi-tetrahedral motif is accommodated by tilting and twisting of the TCNQ linkers; the characteristic "zig-zag" pattern appears when the tilt angles of TCNQ molecules are ordered.[5]

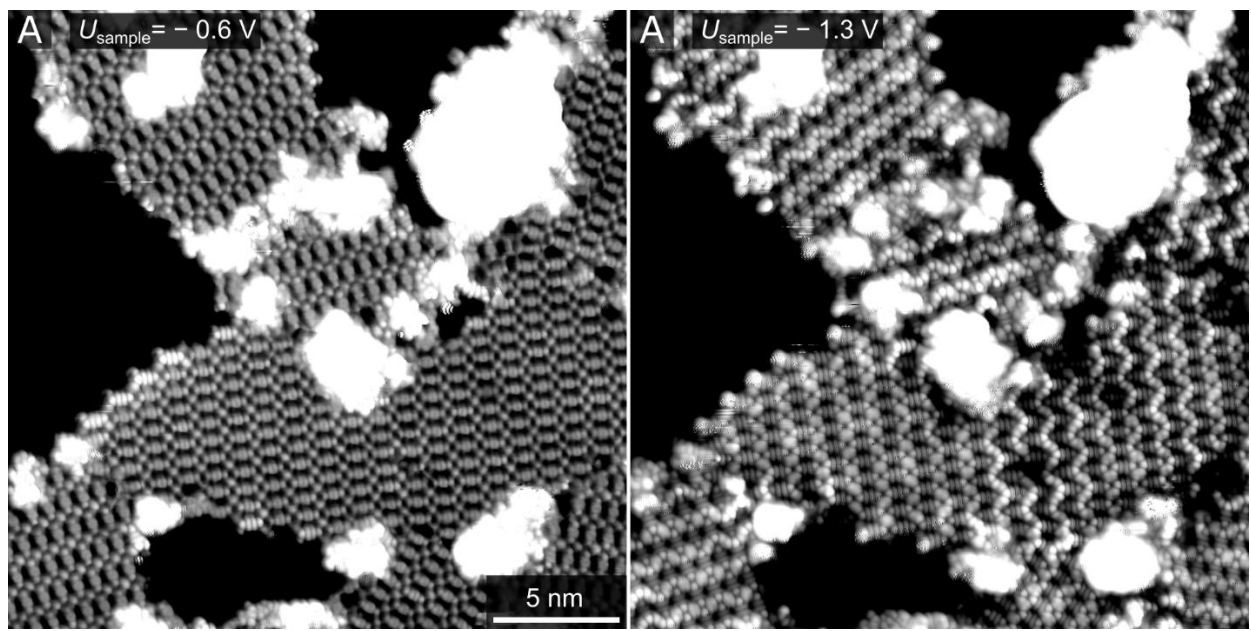

Figure S2: STM images of Fe-TCNQ/gr/Ir(111) taken on the same spot with different bias voltages. (A) At sample bias of –0.6 V, the Fe-TCNQ layer looks homogeneous. (B) At sample bias of –1.3 V, a characteristic "zig-zag" pattern is observed on some areas of the Fe-TCNQ monolayer. Previous literature shows that this is because the Fe-N$_4$ sites prefer quasi-tetrahedral coordination geometry, which can be achieved by an ordered array of differently tilted TCNQ linkers.[5]

## Analysis of Fe-CO bonding using COHP approach

Additional computational analysis was carried out using Crystal Orbital Hamilton Population (COHP) approach to elucidate the contribution of the individual orbitals to the Fe-CO covalent bonding. Figure S3 shows the calculated COHP and integrated COHP (ICOHP) values for CO bonding between the frontier molecular orbitals and the Fe 3d orbitals. Comparison of the CO/Fe-DCA and CO/Fe-TCNQ systems reveals a stronger covalent character of the Fe–CO bond in CO/Fe-DCA, consistent with the higher binding energy. Notably, the ICOHP values associated with LUMO–Fe and LUMO+1–Fe interactions ($\pi$ bonding) are four times lower for Fe-TCNQ, whereas the ICOHP values for HOMO–Fe interactions ($\sigma$ bonding) are only reduced by a factor of two.

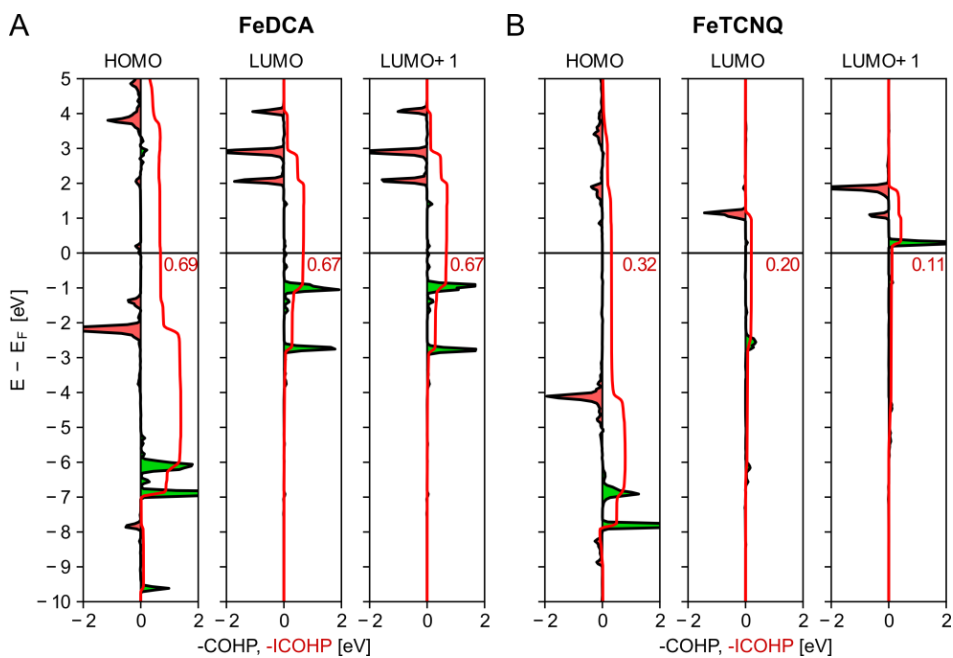

Figure S3: COHP diagrams between the frontier molecular orbitals of CO and Fe 3d orbitals (summed up) for the CO molecule adsorbed on FeDCA (panel A), and FeTCNQ (panel B). The red curve represents a cumulative sum of –COHP (–ICOHP). COHP values stabilizing (destabilizing) a bond are colored in green (red). –COHP values integrated up to Fermi level are written in red.

Next, in Figures S4 and S5 we compare the orbital-resolved contributions to the ICOHP from the individual Fe 3d orbitals (Figure S4 shows this for CO/Fe-DCA, Figure S5 for CO/Fe-TCNQ). The $d_{xy}$ and $d_{x^2-y^2}$ contribution to ICOHP is zero because there is no effective overlap with the CO molecular orbitals. The σ bonding is represented with the $d_{z^2}$ orbital, and π bonding with the $d_{xz}$ and $d_{yz}$ orbitals. Both the σ and π channels show a similar trend: in CO/Fe-DCA, the COHP peaks appear at higher energies and with larger absolute values than in CO/Fe-TCNQ (for example, the $d_{z^2}$ antibonding peak is located at –2.2 eV for CO/Fe-DCA, and –4.1 eV for CO/Fe-TCNQ). This observation is consistent with the basic assumption of the d-band center model: filled electronic states positioned closer to Fermi level (with d-orbital centers at higher energies) hybridize more effectively with the frontier molecular orbitals of CO, resulting in a more pronounced covalent character.

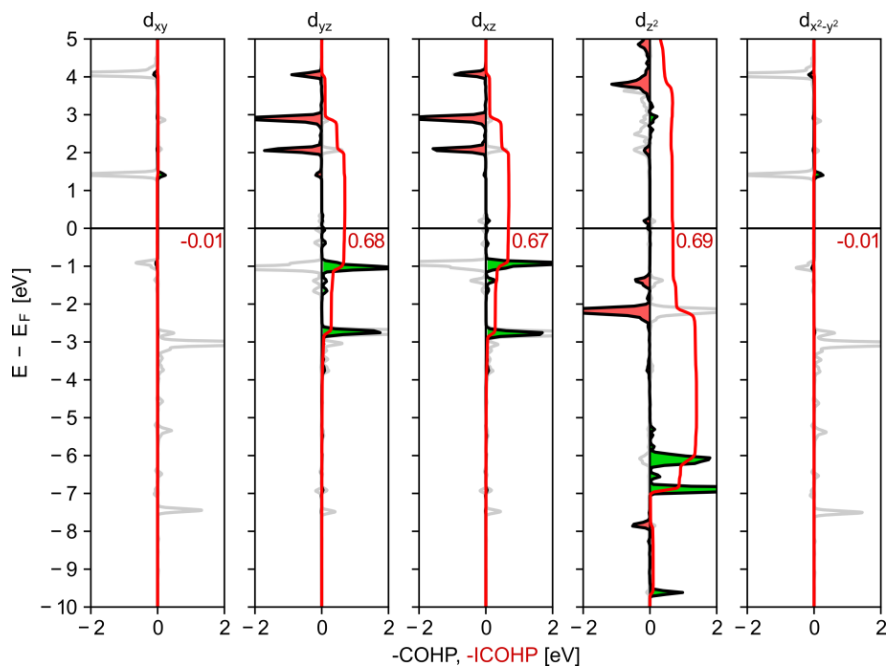

Figure S4: COHP diagrams between the Fe 3d orbitals and the summed frontier molecular orbitals for a CO molecule adsorbed on Fe-DCA. The red curve represents a cumulative sum of –COHP (–ICOHP). COHP values stabilizing (destabilizing) a bond are colored in green (red). –COHP values integrated up to Fermi level are written in red.

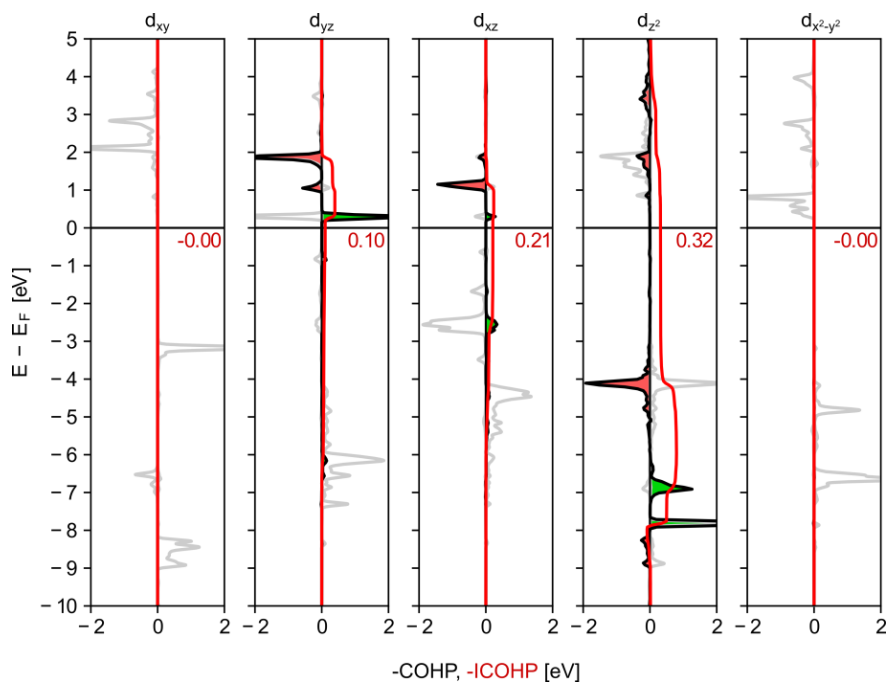

Figure S5: COHP diagrams between the Fe 3d orbitals and the summed frontier molecular orbitals for a CO molecule adsorbed on FeTCNQ. The red curve represents the cumulative sum of –COHP (black curve). The projected density of states is shown in gray. COHP values stabilizing (destabilizing) a bond are colored in green (red). –COHP values integrated up to Fermi level are written in red.

Figure S6 shows the ICOHP values for the $\sigma$ and $\pi$ channels of CO/Fe-DCA and CO/Fe-TCNQ plotted as a function of $\Delta z$ (the lifting of the Fe atom from the $-N_x$ plane). In both systems, the contribution of the $\pi$ channel rises faster than that of the $\sigma$ channel. This indicates that both the Fe $3d_{z^2}$ - CO $5\sigma$ and Fe $3d_{xz/yz}$ - CO $2\pi^*$ interactions are stabilized upon Fe lifting, but the effect is significantly stronger on the Fe $3d_{xz/yz}$ - CO $2\pi^*$.

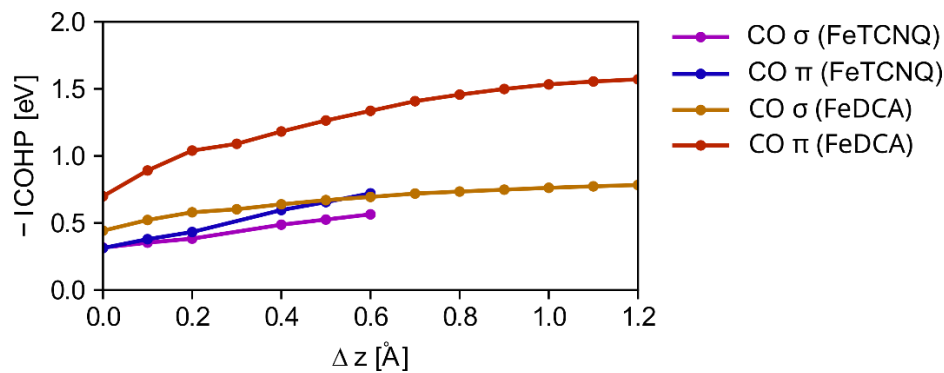

Figure S6: ICOHP values for the bond between the CO molecule and the Fe atom in the FeTCNQ and FeDCA, plotted as a function of the Fe-atom lifting from the $-N_x$ plane.

Figure S7 shows the CO bond lengths and the occupation of the CO LUMO orbital as a function of the Fe atom lifting from the $-N_x$ plane. As expected, both properties show a similar trend, and both are significantly higher in Fe-DCA compared to Fe-TCNQ. This behavior is consistent with the activation of the CO molecule upon adsorption, where increased occupation of the LUMO orbital correlates with stronger Fe-CO adsorption energy and weaker intramolecular C-O bond.

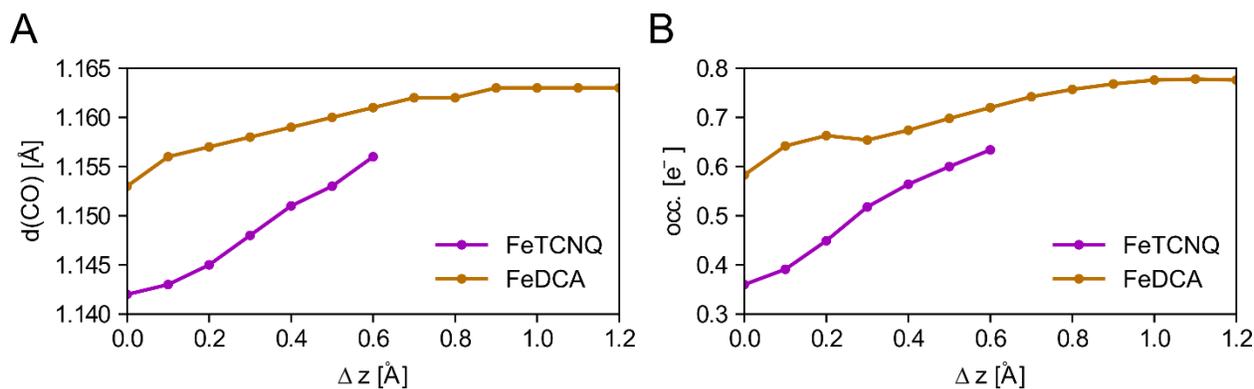

Figure S7: Evolution of the bond length (panel A) and the occupation of the LUMO orbitals (panel B) of the CO adsorbed on the FeTCNQ and FeDCA network with the shift of the Fe atom in the z-axis.

# Analysis of Fe d-orbital center positions upon Fe lifting

Finally, we analyze how the Fe d-orbital center positions evolve upon lifting of the Fe atom from the -$N_x$, in both the Fe-DCA and Fe-TCNQ systems. In Figure S8, we plot the energies of the Fe 3d orbitals centers relative to the Fermi level as a function of the $\Delta z$. Figure S8A,C shows that in Fe-$N_3$ and Fe-$N_4$ sites with fixed planar geometries ($\Delta z = 0$ Å), the Fe d-orbital center positions are very similar; this is consistent with the fact that the CO adsorption energies on these fixed structures are virtually identical, as described in the main text. However, upon lifting of the Fe atom, we observe inconsistent trends in the Fe 3d orbital center positions: In Fe-DCA the d-orbital centers are rising in energy with increasing $\Delta z$ (up to $\Delta z = 1.0$ Å), while in Fe-TCNQ the opposite is happening and the d-orbital center positions are consistently decreasing in energy. This shows that the trends observed in CO binding strengths (shown in Figure 6 in the main text and S8 in the SI) cannot be easily correlated to the position of the d-orbitals centers prior to CO adsorption.

Finally, in Figure S8B,D we plot the position of the Fe 3d orbital centers after CO adsorption as a function of $\Delta z$. Notably, both the absolute positions and qualitative trends are distinct between the two systems (comparison between panels B,D) and also between the CO-free and CO-adsorbed models (comparison between panels A,B and C,D). This behavior is different from the common assumption of the d-band center model applied to extended metallic surfaces, where the adsorption of weak probe molecules at low coverage typically induces only minor changes in the d-band center positions.[18]

In summary, our analysis provides further indications that SAC reactivity descriptors based solely on electronic structure parameters do not fully determine the adsorption properties of "single atoms" embedded within carbon-based materials. Additional parameters including coordination geometry and structural flexibility of the active site must be considered.

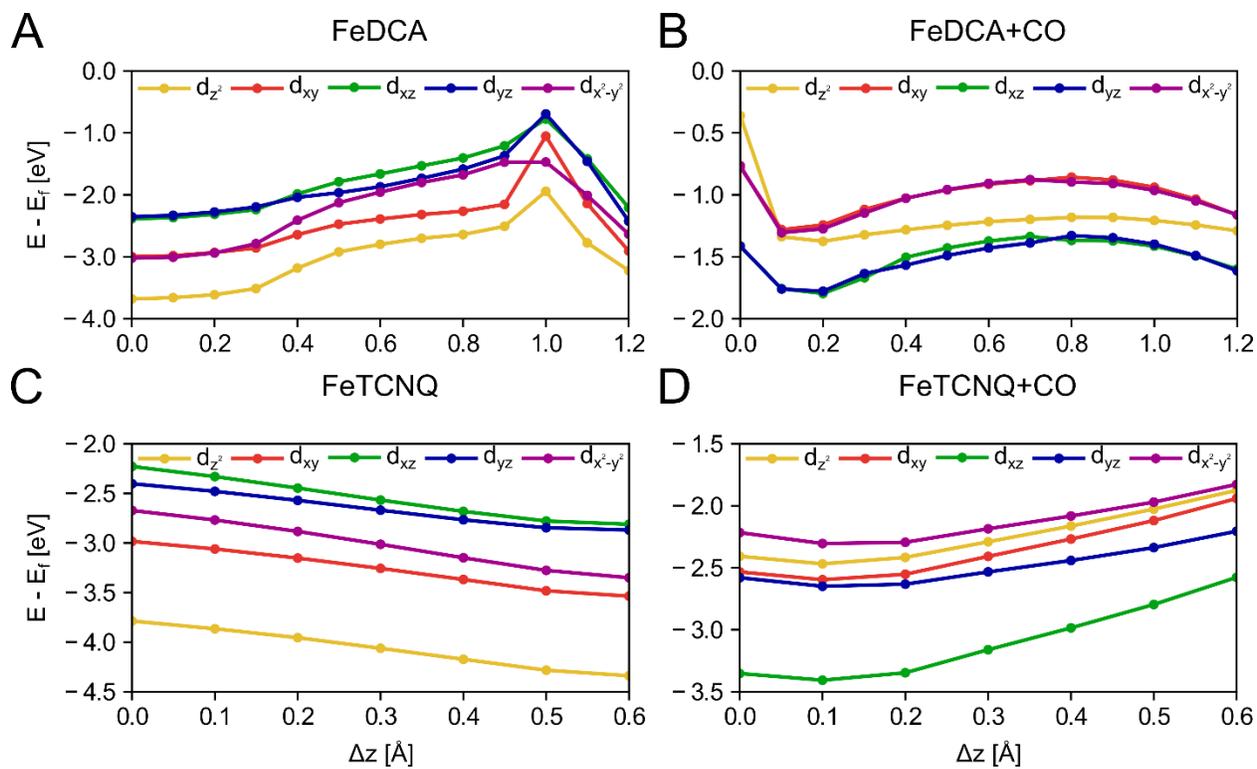

Figure S8: D-band center positions of the Fe atom within Fe-DCA and Fe-TCNQ as a function of the Fe lifting. (A) Fe-DCA without CO, (B) Fe-DCA with CO, (C) Fe-TCNQ without CO, (D) Fe-TCNQ with CO.